\documentclass{80SA}
\usepackage{times}

\title{Correlation Effects on Antiferromagnetism in Fe Pnictides}

\author{Katsunori \textsc{Kubo}$^1$ and Peter \textsc{Thalmeier}$^2$}
\inst{
$^1$Advanced Science Research Center,
Japan Atomic Energy Agency,
Tokai, Ibaraki 319-1195\\
$^2$Max Planck Institute for Chemical Physics of Solids,
01187 Dresden, Germany}

\abst{
  To investigate correlation effects on antiferromagnetic order
  in Fe pnictides,
  we apply a variational Monte Carlo method to a two-orbital model.
  We obtain a small ordered moment
  consistent with experimental observations
  even for a Coulomb interaction
  comparable to the band width.
  Studies of estimation of the Coulomb interaction for Fe pnictides
  suggest values comparable to or slightly smaller than the band width,
  and much larger ordered moments have been obtained by
  the Hartree-Fock approximation for such a large Coulomb interaction.
  Thus, the correlation effect is important for Fe pnictides
  at least quantitatively.
}

\kword{iron pnictides, variational Monte Carlo method, magnetic order}

\begin{document}
\maketitle

The discovery of superconductivity in LaFeAsO$_{1-x}$F$_x$
with a high transition temperature $T_c = 26$~K~\cite{Kamihara}
has stimulated extensive and intensive studies on Fe pnictides.
Superconductivity takes place around the magnetic phase boundaries~\cite{Kamihara,Rotter_dope2,Kotegawa,Chu,Luetkens}
as in high-$T_c$ cuprates.
Such a similarity suggests that
magnetism is playing an important role in the emergence of superconductivity,
and it is highly desirable to unveil the microscopic origin of
magnetism characteristic to Fe pnictides.

To unveil the magnetism in Fe pnictides,
the present authors applied Hartree-Fock approximation~\cite{Kubo:Fe} to a two-orbital model.~\cite{Raghu}
The results are summarized as follows.
The antiferromagnetic order with ordering vector $(\pi,0)$,
in the unfolded Brillouin zone with one Fe ion per unit cell,
is stabilized by the nesting between hole and electron pockets.
This antiferromagnetic state
inevitably accompanies ferro-orbital order,
since the ordering with $(\pi,0)$ breaks the equivalence of
$x$ and $y$ directions, and as a result,
the occupancies of $d_{zx}$ and $d_{yz}$ orbitals become different.
Under such ferro-orbital order,
the lattice should be distorted from a tetragonal to orthorhombic structure
through an electron-lattice interaction.
Even in the antiferromagnetic state,
a band gap does not open at some points in the Brillouin zone
due to multiorbital nature of the bands.
Therefore the system remains metallic in the ordered state.
These results are consistent with experimental observations
of magnetic order with $(\pi, 0)$, lattice distortion,
and metallic conductivity.

However, in our previous Hartree-Fock result,
the ordered moment is large in contradiction with experimental observations,~\cite{Luetkens,Cruz,Zhao,Huang,Kaneko,Rotter_undope,Kitao,Klauss}
if we take a large Coulomb interaction comparable to the band width.
Studies of estimation of the Coulomb interaction for Fe pnictides
suggest values comparable to~\cite{Anisimov} or slightly smaller than~\cite{Nakamura} the band width.
For such large values of Coulomb interactions
correlation effects beyond the Hartree-Fock approximation may be important.
In particular, correlation effects
are expected to reduce the magnitude of the ordered moment.
Indeed, importance of the correlation effects is discussed for a three-orbital model
by using a Gutzwiller approximation.~\cite{Zhou}

In this paper, we investigate correlation effects on magnetism
by applying a variational Monte Carlo (VMC) method to the two-orbital model.
While the VMC method has been applied to a five-orbital model
with a partially-projected Gutzwiller wavefunction,~\cite{Yang}
only Fermi-surface distortion and superconductivity are discussed there.
In the VMC method, we consider a Gutzwiller-projected wavefunction as a variational wavefunction.
We show that this wavefunction contains substantial correlation effects beyond the Hartree-Fock
approximation while this wavefunction is simple enough for numerical calculation.

In the two-orbital model, we consider a square lattice of Fe ions
with $d_{zx}$ and $d_{yz}$ orbitals.~\cite{Raghu,Daghofer}
The model Hamiltonian is given by
\begin{equation}
  \begin{split}
    H=&\sum_{\mib{k},\tau,\tau^{\prime},\sigma}
    \epsilon_{\mib{k} \tau \tau^{\prime}}
    c^{\dagger}_{\mib{k} \tau \sigma}c_{\mib{k} \tau^{\prime} \sigma}
    +U \sum_{i, \tau}
    n_{i \tau \uparrow} n_{i \tau \downarrow}\\
    &+U^{\prime} \sum_{i}
    n_{i x} n_{i y}
    + J \sum_{i,\sigma,\sigma^{\prime}}
    c^{\dagger}_{i x \sigma}
    c^{\dagger}_{i y \sigma^{\prime}}
    c_{i x \sigma^{\prime}}
    c_{i y \sigma}
    \\
    &+ J^{\prime}\sum_{i,\tau \ne \tau^{\prime}}
    c^{\dagger}_{i \tau \uparrow}
    c^{\dagger}_{i \tau \downarrow}
    c_{i \tau^{\prime} \downarrow}
    c_{i \tau^{\prime} \uparrow},
  \end{split}
  \label{eq:H}
\end{equation}
where $c_{i\tau\sigma}$ is the annihilation operator of
the electron at site $i$ with orbital $\tau$
and spin $\sigma$ ($=\uparrow$ or $\downarrow$) and
$c_{\mib{k}\tau\sigma}$ is the Fourier transform of $c_{i\tau\sigma}$.
The orbital indices $\tau=x$ and $y$ represent $d_{zx}$ and $d_{yz}$ orbitals, respectively.
The number operators are defined by
$n_{i \tau \sigma}=c^{\dagger}_{i \tau \sigma} c_{i \tau \sigma}$ and
$n_{i \tau}=\sum_{\sigma}n_{i \tau \sigma}$.
The coupling constants $U$, $U^{\prime}$, $J$, and $J^{\prime}$
denote the intraorbital Coulomb, interorbital Coulomb, exchange,
and pair-hopping interactions, respectively.
The relations $U=U^{\prime}+J+J^{\prime}$ and $J=J^{\prime}$ hold
for the $t_{2g}$ orbitals,~\cite{Tang} and we use them.
We use the hopping parameters proposed
by Raghu~\textit{et al.}~\cite{Raghu} and the coefficients in
the kinetic energy terms are given by
$\epsilon_{\mib{k} xx}=-2t_1\cos k_x-2t_2\cos k_y-4t_3\cos k_x\cos k_y$,
$\epsilon_{\mib{k} yy}=-2t_2\cos k_x-2t_1\cos k_y-4t_3\cos k_x\cos k_y$,
and 
$\epsilon_{\mib{k} xy}=\epsilon_{\mib{k} yx}=-4t_4\sin k_x\sin k_y$,
where
$t_1=-t, t_2=1.3t$, $t_3=t_4=-0.85t$,
and we have set the lattice constant unity.
The band width is $W=12t$.

We consider the variational wave function given by
\begin{equation}
  | \Psi \rangle = P_{\text{G}} | \Phi \rangle
  =\prod_{i \gamma}
  [1-(1-g_{\gamma})|i\gamma\rangle \langle i\gamma |] | \Phi \rangle,
\end{equation}
where $P_{\text{G}}$ is the Gutzwiller projection operator
for onsite density correlation.~\cite{Okabe,Bunemann,Kobayashi2,Kubo:VMC}
$|i\gamma\rangle \langle i\gamma|$ denotes projection onto the state $\gamma$
at site $i$ and $g_{\gamma}$ is the variational parameter
controlling the probability of state $\gamma$.
There are sixteen states at each site in the present two-orbital model.
The Hartree-Fock type wave function $| \Phi \rangle$,
which describes a charge, spin, orbital, and spin-orbital coupled
ordered state, is given by
\begin{equation}
  |\Phi \rangle = \prod_{\mib{k} a \tau \sigma}
  b^{(a) \dagger}_{\mib{k} \tau \sigma} |0 \rangle,
\end{equation}
where $a$ is a band index and $| 0 \rangle$ is the vacuum.
The quasiparticles occupy $N_{\sigma}$ states for each spin $\sigma$
from the lowest quasiparticle energy state,
where $N_{\sigma}$ is the number of electrons with spin $\sigma$.
Here we consider a half-filled case, and we set $N_{\uparrow}=N_{\downarrow}=N$,
where $N$ is the number of the lattice sites.
The quasiparticle states are obtained by diagonalizing the following $4 \times 4$ matrix:
\begin{equation}
  \begin{split}
    \begin{pmatrix}
      \epsilon_{\mib{k} xx} & \epsilon_{\mib{k} xy} & 0                           &  0                       \\
      \epsilon_{\mib{k} yx} & \epsilon_{\mib{k} yy} & 0                           &  0                       \\
      0                   & 0                    & \epsilon_{\mib{k}+\mib{Q} xx} & \epsilon_{\mib{k}+\mib{Q} xy}\\
      0                   & 0                    & \epsilon_{\mib{k}+\mib{Q} yx} & \epsilon_{\mib{k}+\mib{Q} yy}
    \end{pmatrix}\\
    -
    \begin{pmatrix}
      \Delta_{x \sigma}        & 0                      & \Delta_{x \sigma \mib{Q}} &  0                     \\
      0                      & \Delta_{y \sigma}        & 0                       & \Delta_{y \sigma \mib{Q}}\\
      \Delta_{x \sigma \mib{Q}} &  0                     & \Delta_{x \sigma}        & 0                      \\
      0                      & \Delta_{y \sigma \mib{Q}} & 0                       & \Delta_{y \sigma}
    \end{pmatrix}
  \end{split},
\end{equation}
where $\mib{Q}=(\pi,0)$ is the ordering vector.
The quasiparticle gap in the ordered state is given by
\begin{equation}
    \Delta_{\tau \sigma}=
     \Delta_{\text{o}}(\delta_{\tau x}-\delta_{\tau y})
    +\Delta_{\text{so}}(\delta_{\sigma \uparrow}-\delta_{\sigma \downarrow})(\delta_{\tau x}-\delta_{\tau y}),
  \label{eq:gap}
\end{equation}
\begin{equation}
  \begin{split}
    \Delta_{\tau \sigma \mib{Q}}=&\Delta_{\text{c} \mib{Q}}+\Delta_{\text{s} \mib{Q}}(\delta_{\sigma \uparrow}-\delta_{\sigma \downarrow})
    +\Delta_{\text{o} \mib{Q}}(\delta_{\tau x}-\delta_{\tau y})\\
    &+\Delta_{\text{so} \mib{Q}}(\delta_{\sigma \uparrow}-\delta_{\sigma \downarrow})(\delta_{\tau x}-\delta_{\tau y}),
  \end{split}
  \label{eq:gapQ}
\end{equation}
where $\Delta_{\text{o}}$ and $\Delta_{\text{so}}$
denote the gaps for uniform orbital and spin-orbital ordered states, respectively.
$\Delta_{\text{c} \mib{Q}}$, $\Delta_{\text{s} \mib{Q}}$, $\Delta_{\text{o} \mib{Q}}$, and $\Delta_{\text{so} \mib{Q}}$
denote the gaps for antiferro-ordered states of charge, spin, orbital, and spin-orbital, respectively.
We also take them as variational parameters.

For this variational wavefunction, we evaluate energy by the Monte Carlo method,
and optimize variational parameters to find the state which has the lowest energy.
We set all $\Delta_{\tau \sigma}$ and $\Delta_{\tau \sigma \mib{Q}}$ zero
to evaluate energy of the paramagnetic state, that is,
we optimize only Gutzwiller parameters $g_{\gamma}$.
For the antiferromagnetic state, we also vary $\Delta_{\text{o}}$, $\Delta_{\text{s} \mib{Q}}$,
and $\Delta_{\text{so} \mib{Q}}$.
We also evaluated energy by varying all $\Delta_{\tau \sigma}$ and $\Delta_{\tau \sigma \mib{Q}}$
for some values of $U$,
but we could not find a solution which has lower energy than the antiferromagnetic state.
The calculations are done for an $8 \times 8$ lattice
with an antiperiodic boundary condition for both directions.

\begin{figure}
\begin{center}
\includegraphics[width=\linewidth]{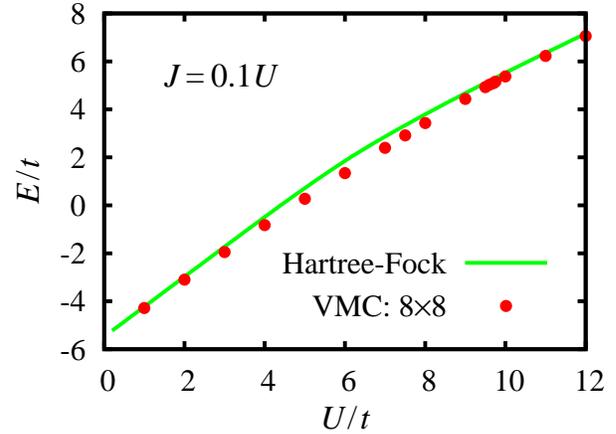}
\end{center}
\caption{(Color online)
  Energy $E$ as functions of the Coulomb interaction with $J=0.1U$
  obtained with the Hartree-Fock approximation~\cite{Kubo:Fe}
  and the VMC method.
}
\label{E_and_EHF}
\end{figure}
Figure~\ref{E_and_EHF} shows energy as functions of $U$ obtained with the Hartree-Fock approximation~\cite{Kubo:Fe}
and the present VMC method.
The energy is lowered by the correlation effects beyond the Hartree-Fock approximation.

\begin{figure}
\begin{center}
\includegraphics[width=\linewidth]{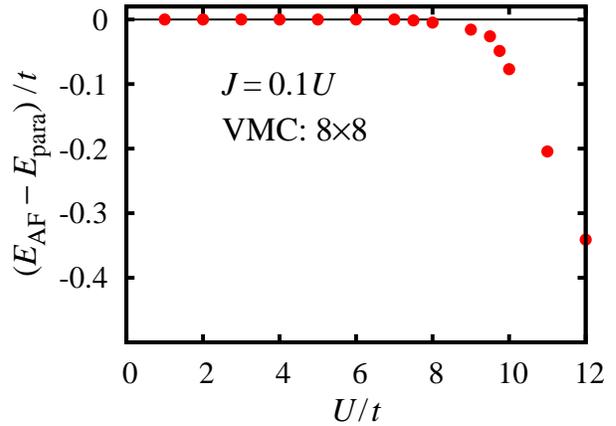}
\end{center}
\caption{(Color online)
  Energy $E_\text{AF}$ of the antiferromagnetic ground state measured from
  energy $E_\text{para}$ of the paramagnetic state as a function of $U$
  with $J=0.1U$.
}
\label{EAF_Epara}
\end{figure}
Figure~\ref{EAF_Epara} shows the ground state energy as a function of $U$
measured from that of the paramagnetic state.
The transition from the paramagnetic state to the antiferromagnetic state
occurs at $U \gtrsim 7t$.
If the energy difference is proportional to $U^2$ around the transition it is of second order,
and if the energy difference is proportional to $U$ the transition is first order.
However, it is difficult to distinguish a second order transition
from a weak first order transition as is obtained by the Hartree-Fock approximation~\cite{Kubo:Fe}
from the present results
due to numerical accuracy.

\begin{figure}
\begin{center}
\includegraphics[width=\linewidth]{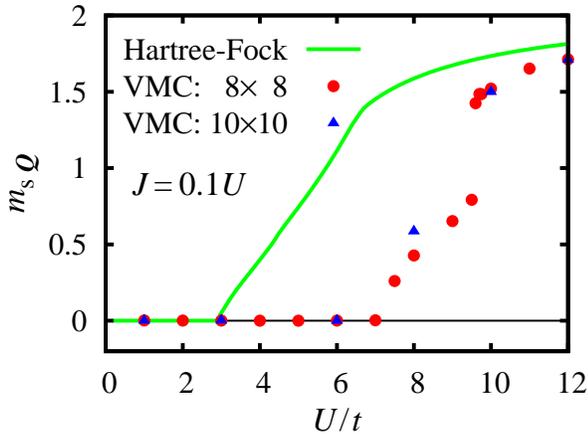}
\end{center}
\caption{(Color online)
  Ordered moment $m_{\text{s} \mib{Q}}$ in the antiferromagnetic ground state
  as functions of $U$ with $J=0.1U$
  obtained with the Hartree-Fock approximation~\cite{Kubo:Fe} and the VMC method.
  Note that if a fully polarized state is realized, it has $m_{\text{s} \mib{Q}}=2$.
}
\label{order_param}
\end{figure}
Figure~\ref{order_param} shows the ordered magnetic moment $m_{\text{s} \mib{Q}}$
evaluated for the optimized wavefunction.
$m_{\text{s} \mib{Q}}$ is defined as
\begin{equation}
  m_{\text{s} \mib{Q}}=\frac{1}{N}\sum_{i \tau}\text{e}^{\text{i}\mib{Q}\cdot\mib{r}_i}
  \langle n_{i \tau \uparrow}- n_{i \tau \downarrow} \rangle,
\end{equation}
where $\mib{r}_i$ denotes the position of site $i$
and $\langle \cdots \rangle$ represents the expectation value.
To check the finite size effect of the model,
we also show the results for a $10 \times 10$ lattice.
The finite size effect on $m_{\text{s} \mib{Q}}$ is weak
in particular for the large $m_{\text{s} \mib{Q}}$ region.
In the result of the Hartree-Fock approximation,
there is a small but finite jump in $m_{\text{s} \mib{Q}}$,~\cite{Kubo:Fe}
while it is invisible in the scale of Fig.~\ref{order_param}.
For the results of the VMC, as in the energy difference,
it is difficult to determine whether the transition
is first order or second order.
If it is a first order transition,
the jump in the magnetic moment at the transition is very small.
The ordered moment is not large for $U \lesssim 9t$.
By comparing the results by the Hartree-Fock approximation
and the present VMC results,
we conclude that the correlation effect strongly reduces the value of the ordered
moment and such an effect is important for Fe pnictides.

Around $U=9.6t$,
we find another phase transition within the antiferromagnetic phase.
This phase transition is of first order
and it is probably a metal to insulator transition,
since the energy gain by the kinetic energy is reduced at $U \gtrsim 9.6t$ (not shown).

We have also searched for a ferro-orbital ordered state without antiferromagnetic order,
that is, the gap parameters are set zero except for $\Delta_{\text{o}}$,
but we could not find such a state as a ground state.
In the antiferromagnetic state,
the order parameter $m_{\text{o}}=(1/N)\sum_{i \sigma}\langle n_{i x \sigma}-n_{i y \sigma} \rangle$
for the ferro-orbital order becomes also finite due to symmetry lowering.
However, the values are too small and we cannot determine $m_{\text{o}}$ confidently
due to our numerical accuracy.

To summarize, we have applied the variational Monte Carlo method to a two-orbital model
to investigate correlation effects.
Then, we have found that the ordered moment in the antiferromagnetic state is
strongly suppressed by the correlation effect.
Thus, to obtain a small ordered moment as in experimental observations, for $U \lesssim W$,
we should take correlation effect into account properly.

\end{document}